\documentclass[runningheads]{llncs}
\usepackage[T1]{fontenc}
\usepackage{graphicx}
\usepackage{CJKutf8}
\usepackage{amsmath} 
\usepackage[T1]{fontenc}
\usepackage[utf8]{inputenc}
\usepackage{subcaption}
\usepackage{algorithm} 
\usepackage{algpseudocode} 
\usepackage{latexsym}
\usepackage{bbm}
\usepackage{dsfont}
\usepackage{threeparttable}
\usepackage{enumerate}
\usepackage[inline]{enumitem}
\usepackage{float}
\usepackage{lipsum} 
\usepackage{float}
\usepackage{setspace}
\usepackage{booktabs} 
\usepackage[english]{babel}
\usepackage{multirow}
\usepackage{xspace}
\usepackage{xcolor}
\usepackage{microtype}
\usepackage{graphicx}

\newcommand{\header}[1]{\vspace*{1mm}\noindent\textbf{#1.}}
\newcommand{\eg}{\emph{e.g.,}\xspace}
\newcommand{\ie}{\emph{i.e.,}\xspace}

\AtBeginDocument{%
  }

\begin{document}
\title{LLM-based Listwise Reranking under the Effect of Positional Bias}

\author{
Jingfen Qiao\inst{1}\thanks{Corresponding author} \and
Jin Huang\inst{2} \and
Xinyu Ma\inst{3} \and
Shuaiqiang Wang\inst{3} \and
Dawei Yin\inst{3} \and
Evangelos Kanoulas\inst{1} \and
Andrew Yates\inst{4} 
}
\authorrunning{J. Qiao et al.}

\institute{
University of Amsterdam, Amsterdam, The Netherlands\\
\email{\{j.qiao, e.kanoulas\}@uva.nl}
\and
University of Cambridge, United Kingdom \\
\and
Baidu Inc., China
\and
Johns Hopkins University, United States\\
\email{andrew.yates@jhu.edu}
}

\maketitle              
\begin{abstract}
LLM-based listwise passage reranking has attracted attention for its effectiveness in ranking candidate passages. However, these models suffer from positional bias, where passages positioned towards the end of the input are less likely to be moved to top positions in the ranking.  We hypothesize that there are two primary sources of positional bias: (1) architectural bias inherent in LLMs and (2) the imbalanced positioning of relevant documents. To address this, we propose DebiasFirst, a method that integrates positional calibration and position-aware data augmentation during fine-tuning. Positional calibration uses inverse propensity scoring to adjust for positional bias by re-weighting the contributions of different positions in the loss function when training. Position-aware augmentation augments training data to ensure that each passage appears equally across varied positions in the input list. This approach markedly enhances both effectiveness and robustness to the original ranking across diverse first-stage retrievers, reducing the dependence of NDCG@10 performance on the position of relevant documents. DebiasFirst also complements the inference-stage debiasing methods, offering a practical solution for mitigating positional bias in reranking.
\end{abstract}

\section{Introduction}

Large language models (LLMs) have received increased attention for their applications in information retrieval (IR)~\cite{zhu2023large,zhai2024large}. Listwise passage reranking is a core application of LLMs in IR, aiming to rank a list of candidate passages. Sun et al.~\cite{sun2023chatgptgoodsearchinvestigating} found that, with proper instructions, GPT-4 can deliver competitive, even superior results to state-of-the-art supervised methods for listwise reranking. Subsequent studies proposed to fine-tune open-source LLMs to perform listwise reranking, thereby improving the efficacy of this task~\cite{pradeep2023rankvicuna,pradeep2023rankzephyr,reddy2024firstfasterimprovedlistwise,yoon2024listt5,Ruiyang_selfcalibrated_2025}.

Recent studies have highlighted \emph{positional bias}, where LLMs prioritize content based on its position within the given context~\cite{hofstätter2021mitigatingpositionbiastransformer,wang2024eliminating,wang2023large,liu2024lost,chen2024attentionlargelanguagemodels,shi2025judgingjudgessystematicstudy,chen2024llmsbiasedevaluatorsbiased}. Several forms of positional bias have been observed, \eg prompt order effects~\cite{wang2023large}, where certain orders outperform others, and the ``lost in the middle'' phenomenon~\cite{liu2024lost}, where performance degrades when relevant information is in the middle of long contexts. Tang et al.~\cite{tang2023found} identified positional bias that
depends on the pairwise positions of items in the ranking list.
To investigate how input position affects reranking performance, we evaluate a widely used reranking model by changing the position of the relevant passage within its input. We observed that ranking performance is reduced when a relevant passage is not positioned at the beginning of the input, as detailed in Section 6. This positional bias makes reranking performance fundamentally unreliable and overly dependent on ordering.

To address positional bias, recent studies have proposed methods targeting the inference and fine-tuning stages of LLM-based reranking methods, respectively. At the inference stage, most approaches mitigate positional bias by aggregating output rankings generated from different input orders~\cite{tang2023found,hou2024largelanguagemodelszeroshot,zeng2024llmrankfusionmitigatingintrinsicinconsistency}. For example, PermSC~\cite{tang2023found} aggregates rerankings across various permutations and derives a central ranking that minimizes positional bias by selecting the one closest to all permutations under distance metrics. Alternatively, ListT5~\cite{yoon2024listt5} generates a sorted list of input passages in increasing order of relevance, progressively eliminating irrelevant passages to deduce the most relevant passages, thereby mitigating positional bias. While effective, these approaches require additional memory~\cite{izacard2020leveraging} or multiple inference runs~\cite{tang2023found}. Apart from the inference stage, some studies have attempted to eliminate positional bias in listwise reranking by introducing random shuffling augmentations of fine-tuning data~\cite{pradeep2023rankvicuna,pradeep2023rankzephyr}. Yet, these models still suffer from positional bias: our experiments show that ranking performance is reduced when relevant passages are not in the initial position. Motivated by this, we propose a method that improves robustness to variations in input order at the fine-tuning stage, ensuring more stable performance across all positions.

To better reduce positional bias in LLM-based listwise reranking, we first conduct a causal analysis through hypothesis testing, assuming bias arises from (1) architectural bias inherent in LLMs and (2) the imbalanced positioning of relevant documents. Building on this, we propose a debiasing method for LLM-based listwise reranking, consisting of two components: positional calibration using inverse propensity scoring (IPS) and position-aware augmentation. IPS-based positional calibration adjusts for positional bias by re-weighting the contributions of different positions in the loss function for reranking. Position-aware augmentation augments training data to ensure that each passage appears equally across different positions of the input list. Together, these techniques substantially improve robustness to input order, reducing the dependence of NDCG@10 performance on the position of relevant documents. This reduction leads to improved ranking performance across both in-domain and out-of-domain datasets, achieving a 2\%-4\% increase in average NDCG@10 depending on the setting.
\vspace{-2mm}
\section{Preliminary}
\vspace{-2mm}
Given an instruction prompt $X$ that includes a query \(q\) and a list of candidate passages $\{x_i | 1 \leq i \leq k\}$, listwise rerankers aim to rerank these passages simultaneously, ensuring that those most relevant to query \(q\) appear higher in the reranked list. 
The position of passage \(x_i\) in the reranking permutation is denoted as \(\hat{\pi}_q(x_{i})\), which is determined by the relevance score $f_\theta(x_i)$ predicted by the LLM-based method.
For simplicity, we omit the superscript $q$ when the query is clear from the context.
The true reranking permutation comes in the form of sequence \(y = [y_1] > [y_2] ... > [y_k]\), where $y_i$
denotes the identifier of the $i$-th most relevant document.
Recent work~\cite{pradeep2023rankvicuna,pradeep2023rankzephyr} has proposed fine-tuning LLMs as rerankers with a language modeling (LM) objective, minimizing the prediction error in predicting correct document identifiers in the generation sequence:
\begin{equation}
    \mathcal{L}_\text{LM} = - \sum_{i=1}^{|y|} \log(P_{\theta}(y_i|X, y_{<i})),
    \label{eq:1}
\vspace{-3mm}
\end{equation}
where $P_\theta$ denotes the conditional probability of predicting the target $y_i$ based on the prompt $X$ and the preceding identifiers $y_{<i}$.

Such LLM reranking methods lack efficiency as they generate the reranking permutation in the form of an ordered sequence of candidate passage identifiers. 
To improve efficiency, First~\cite{reddy2024firstfasterimprovedlistwise} proposes leveraging the output logits of the first generated identifier to directly derive a ranked ordering of the input passages.
Specifically, $f_\theta(x_i)$ is the output vocabulary logit for the passage identifier of passage $x_i$ during the first token generation in the First model.
Building on this, First formulates its training objective as follows:
\begin{equation}
    \mathcal{L}_\text{Rank} = \sum_{\pi(x_i) < \pi(x_j)} \frac{1}{i+j} \delta(f_\theta(x_i), f_\theta(x_j)).
    \label{eq:first-ltr}
\vspace{-2mm}
\end{equation}

\noindent
A logistic loss function $\delta(\cdot)=\log(1+\exp(\cdot))$ is applied to measure the relative difference in the predicted relevance scores between two passages. 
Here, the weight $1/(i+j)$ assigns greater weights to higher ranked passages, reducing the risk of misranking those that are more relevant.

Drawing from the success of the language modeling objective in listwise reranking, First combines $\mathcal{L}_\text{Rank}$ and $\mathcal{L}_\text{LM}$ into the final joint loss for fine-tuning the LLM parameters~$\theta$: 
\begin{equation}
    \mathcal{L}_\text{First} = \lambda\mathcal{L}_\text{Rank} + \mathcal{L}_\text{LM},
\vspace{-2mm}
\end{equation}
where $\lambda$ controls the relative importance of two losses.

\section{Causal Analysis of Positional Bias}

\begin{figure}
    \vspace{-3mm}
    \centering
    \includegraphics[width=0.3\linewidth, trim=0 15pt 0 8pt]{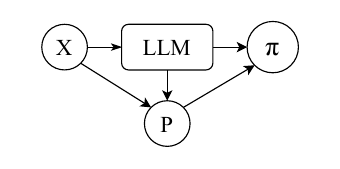}
    \caption{A causal directed acyclic graph illustrating the relationships between input (X), LLM, positional bias (P), and the resulting output ($\pi$) in listwise reranking.}
    \label{fig:causal_bias}
\vspace{-3mm}
\end{figure}

Before introducing our method to mitigate positional bias, we first analyze the causes of positional bias associated with LLMs in listwise passages reranking.
Inspired by literature~\cite{su2024roformer,sun2023chatgptgoodsearchinvestigating,tang2023found,xiao2023efficient}, we demonstrate the causal relationships between data (X), LLM model (LLM), positional bias (P), and the resulting reranking permutation ($\pi$) in Figure~\ref{fig:causal_bias} and hypothesize the following two primary sources of positional bias.

\noindent
\textbf{Bias arising from the architecture of LLM} (LLM $\rightarrow$ P).
LLMs commonly use rotary position embedding (RoPE)~\cite{su2024roformer}, a type of relative positional encoding that encodes absolute positional information using rotation matrix, thus enhancing their ability to capture and manipulate long-term dependencies between tokens in sentences.
However, architecture often diminishes focus on tokens located further from the query, as the scale of attention decays over distance~\cite{chen2024hope}. 
Moreover, Xiao et al.~\cite{xiao2023efficient} observed that earlier tokens in a sequence receive disproportionately high attention logits, even when they are not semantically important. 

Additionally, LLM-based rerankers often face generation failure, where LLMs fail to predict passage identifiers. 
A common solution is to append unpredicted passages in their original input order from the first stage retriever to the end of LLMs' output~\cite{sun2023chatgptgoodsearchinvestigating}. This may exacerbate positional bias as it preserves the initial order of passages in the reranking output.

\noindent
\textbf{Bias arising from imbalanced relevance distribution across input positions} (X~$\rightarrow$~P).
Training data used for fine-tuning LLMs often exhibits an unequal distribution of relevant documents across different positions in the input sequence, as shown on the left of Figure \ref{fig: aug}.
In general, we observe that passages initially positioned higher in the input are significantly more likely to be relevant in the MS MARCO training set, as evidenced by the higher frequency of these passages occupying higher positions in the true reranking permutation.
This imbalance can heighten the sensitivity of LLMs to positional cues, thereby exacerbating bias toward early passages in the input during the fine-tuning process.

Together, the imbalanced relevance distribution in the training data (X) and the positional bias inherent in the LLM architecture (LLM) contribute to a great emphasis on early-positioned passages (P) when fine-tuned LLMs are used to generate reranking permutations $\hat{\pi}$.

\section{Method}

\begin{figure*}
    \centering
    \includegraphics[width=1.0\linewidth]{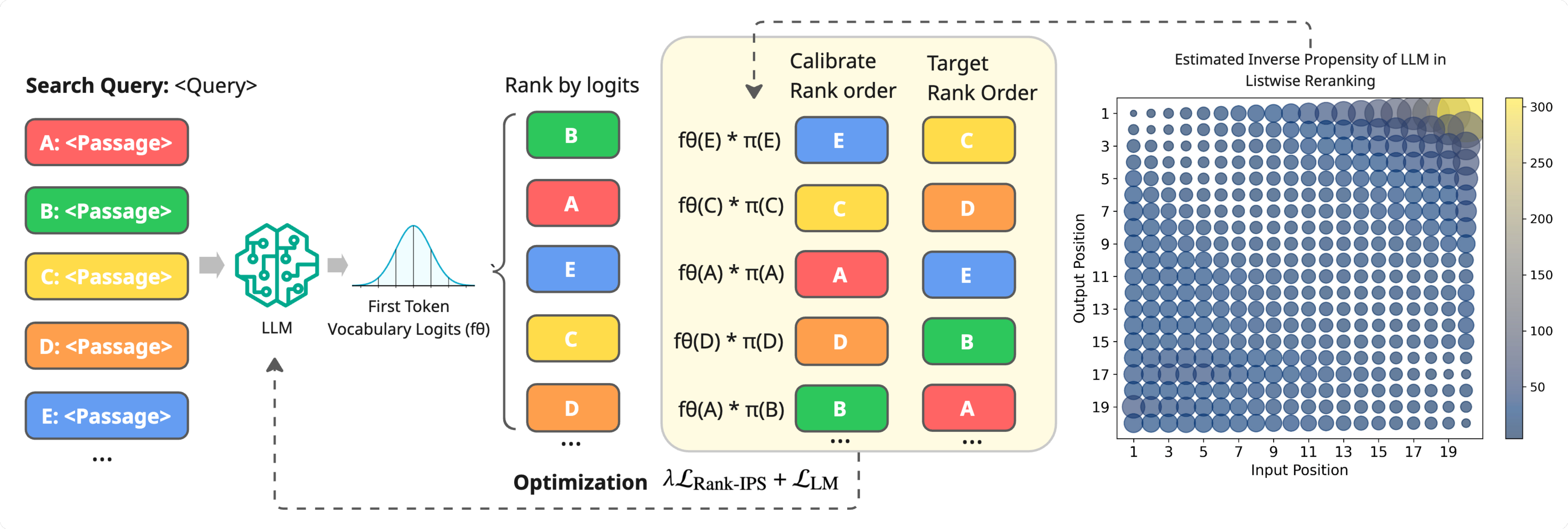}
    \caption{Overview of the proposed positional calibration using IPS. Each document’s relevance score $f_\theta(x_i)$ is calibrated by multiplying it with estimated inverse propensity values \(\pi_q(x_{i})\) to account for positional bias. The heatmap on the right visualizes the estimated inverse propensities across input and output positions.}
    \label{fig: ips_overview}
\end{figure*}

Our method DebiasFirst mitigates positional bias in fine-tuned LLMs for listwise reranking through two components: positional calibration and position-aware augmentation. Positional calibration operates at the loss-function level, employing inverse propensity scoring (IPS) to adjust the loss contribution of underrepresented or overrepresented input positions. Position-aware augmentation (Pos-Aug) enhances the robustness of the model by ensuring that each passage appears equally across various positions in the input list.

\subsection{Positional Calibration through IPS}

To reduce the positional bias that causes varying attention across different input positions, we want to balance the contribution of each passage by adjusting the influence of transitions between input and reranking positions based on their frequency, as illustrated in Figure~\ref{fig: ips_overview}.

Formally, given a passage $x_i$ and its true relevant position $\pi(x_i)$,
let $\omega_{i,\pi(x_i)}$ denote the influence of the transition between the input and reranking positions, referred to as propensity, which quantifies positional bias in LLM fine-tuning.
Building on the learning-to-rank objective in the First model (See Eq.~(\ref{eq:first-ltr})) and the idea of IPS,
we adjust the loss for a pair of passages \((x_i, x_j)\) by inversely weighting it with the product of their propensities $\omega_{i,\pi(x_i)}$ and $\omega_{j,\pi(x_j)}$:

\begin{equation}
    \mathcal{L}_\text{Rank-IPS} = \sum_{\pi(x_i) < \pi(x_j)} \frac{\delta(f_\theta(x_i), f_\theta(x_j))}{(i+j)\cdot  \omega_{i,\pi(x_i)} \cdot  \omega_{j,\pi(x_j)}}.
    \label{eq:4}
\end{equation}

If certain transitions (\ie from certain input positions to certain reranking positions) are overrepresented with high propensities, $\mathcal{L}_\text{Rank-IPS}$ assigns low weights to these transitions, thereby reducing the influence of positional bias on the training process. 
Conversely, $\mathcal{L}_\text{Rank-IPS}$ assigns a high weight to transitions that are frequently underrepresented with low propensities.

\header{Positional Bias Estimation} 
$\mathcal{L}_\text{Rank-IPS}$ requires accurate propensities to remove the effect of positional bias, which we estimate by adoptiong a randomization strategy inspired by previous work~\cite{Ai_2018}. We shuffle passages \(n\) times for each query to augment the original prompt set $\mathcal{X}$, denoted as $\widetilde{\mathcal{X}}$. 

\begin{figure}
    \centering
    \begin{minipage}{.48\linewidth}
        \centering
        \begin{minipage}{.48\linewidth}
            \centering
            \includegraphics[width=\linewidth, trim=0 15pt 0 0]{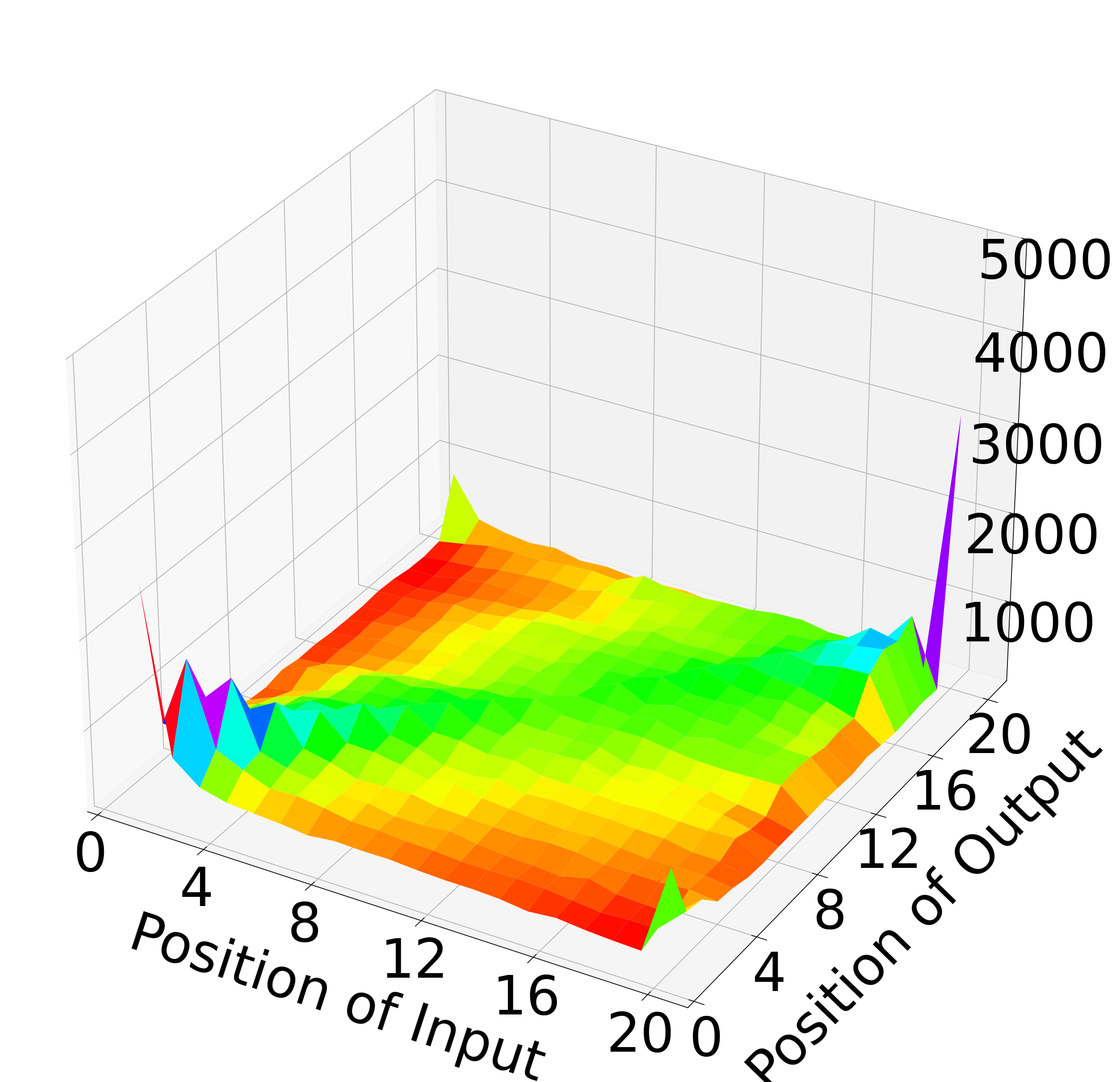}
            \caption*{(a) Original}
            \label{fig:original}
        \end{minipage}
        \hfill
        \begin{minipage}{.48\linewidth}
            \centering
            \includegraphics[width=\linewidth, trim=0 15pt 0 0]{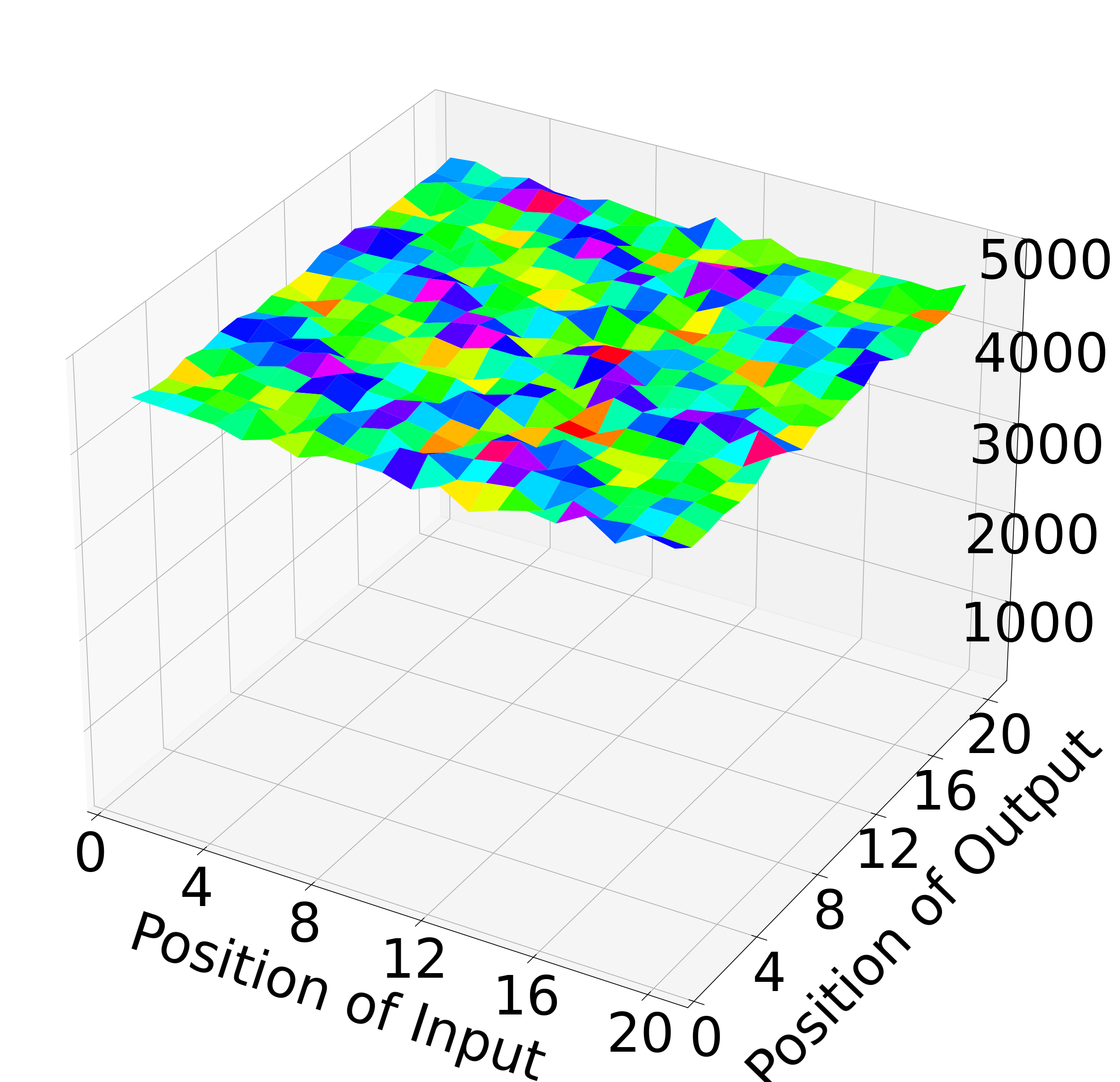}
            \caption*{(b) Using Pos-Aug}
            \label{fig:aug}
        \end{minipage}
        \caption{Number of passages (z) with input positions (x) and  true reranking positions~(y).}
    \label{fig: aug}
    \end{minipage}
    \hfill
    \begin{minipage}{.48\linewidth}
        \centering
        \includegraphics[width=1\linewidth, trim=0pt 30pt 0 0]{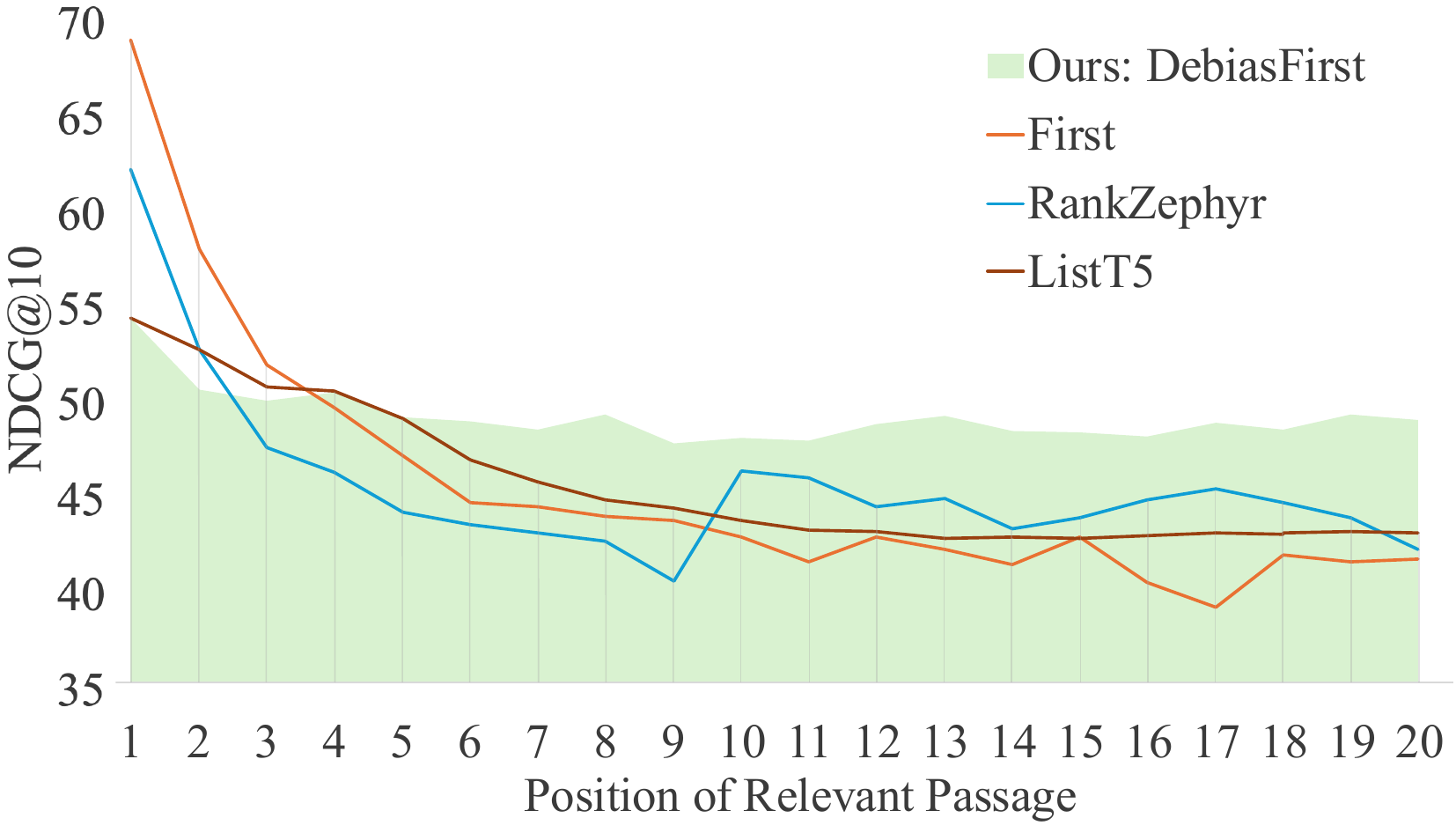}
        \caption{Performance of LLM-based reranking methods when changing the position of the relevant passages within their input on MS MARCO (dev).}
        \label{fig: baselines}
    \end{minipage}
\end{figure}

Given input position $i$ and any position output position $\overline{\pi}$, the propensity $\omega_{i, \overline{\pi}}$ can be estimated as the fraction of all transitions assigned to the reranking position $\overline{\pi}$ from input position $i$ in the modified prompt set $\widetilde{\mathcal{X}}$:

\begin{equation}
\omega_{i,\overline{\pi}} = \frac{\sum_{(q, i) \in \widetilde{\mathcal{X}}} \mathbbm{1}_{\hat{\pi}_{q} (x_{i})=\overline{\pi}}}{|Q| \cdot k \cdot n },
\end{equation}

where $|Q|$ is the number of queries and
\(|Q| \cdot k \cdot n\) is the total number of passages observed in all the permutations in $\widetilde{\mathcal{X}}$. 

\begin{enumerate}
    \item \textbf{Unbiased Random Shuffling}: Initially, the passages for each query are randomly shuffled using the Fisher-Yates shuffling algorithm~\cite{simoes1964algorithm}. It produces an unbiased permutation, ensuring that every passage has an equal probability of appearing in any position~\cite{eberl2016fisher}.

    \item \textbf{Grouping and Rotation}: The shuffled passages are then divided into \(n\) groups. By rotating these groups in order, we generate different \(n\) permutations of the passage order, ensuring a balanced distribution of relevant passages across both input and output positions. 
\end{enumerate}

\subsection{Position-Aware Augmentation}
To further mitigate positional bias arising from the imbalanced distribution of relevant passages in the training data, we implement a position-aware augmentation (Pos-Aug) strategy to enhance training data. This augmentation strategy ensures each passage appears in a wide variety of positions across multiple training instances. By exposing the model to diverse positional contexts, we reduce the likelihood of the model overfitting to specific positional patterns. Our proposed Pos-Aug involves two primary steps:

Figure~\ref{fig: aug} illustrates the movement of passages in the training data from their input positions to reranking positions, showing the frequency of each transition before and after applying position-aware augmentation. We clearly observe varied passage orders in the augmented training data after applying Pos-Aug, which helps reduce the influence of positional bias when fine-tuning on such data.

Finally, by combining both positional calibration with IPS and the position-aware augmentation, we present DebiasFirst, a comprehensive method to mitigate positional bias in LLM-based listwise passages reranking.
Building on previous findings regarding the benefits of jointly fine-tuning with a language modeling objective~\cite{reddy2024firstfasterimprovedlistwise}, our final training objective is defined below:

\begin{equation}
    \mathcal{L}_\textrm{DebiasFirst} = \lambda\mathcal{L}_\text{Rank-IPS} + \mathcal{L}_\text{LM}.
    \label{eq:5}
\end{equation}

\noindent
The LLM parameters $\theta$ are updated by optimizing $\mathcal{L}_\textrm{DebiasFirst}$ on data augmented using our position-aware augmentation strategy.

\section{Experimental Setup}

\header{Dataset and evaluation} Our study utilizes the same training dataset as in previous works by~\cite{reddy2024firstfasterimprovedlistwise} and~\cite{pradeep2023rankvicuna,pradeep2023rankzephyr} to ensure consistency for comparative analysis. This dataset, which comprises 40K instances labeled by GPT-4 and originating from~\cite{pradeep2023rankzephyr}, was created using 5K queries from MS MARCO~\cite{bajaj2016ms}. Our positional-aware augmentation techniques are implemented on this training data. The study evaluates model performance across both in-domain datasets (MS MARCO Dev, TREC DL 2019 and 2020) and an out-of-domain dataset (BEIR).  Three different first-stage retrievers (Contriever~\cite{izacard2021contriever}, Splade++~\cite{splade++}, BM25) are used. All evaluations use NDCG@10 as the relevance metric.

\header{Baselines} We compared our method against two categories of baselines: (1) fine-tuned LLMs for listwise passages reranking and (2) the inference-stage debiasing approach. In terms of fine-tuned LLMs, we evaluated our approach against RankZephyr~\cite{pradeep2023rankzephyr}, First~\cite{reddy2024firstfasterimprovedlistwise}, and ListT5~\cite{yoon2024listt5}. 
RankZephyr uses random augmentation to mitigate positional bias, while ListT5 eliminates positional bias by jointly considering the relevance of multiple candidate passages at both the training and inference stages. First~\cite{reddy2024firstfasterimprovedlistwise} does not incorporate any positional bias mitigation techniques at either the training or inference stage.  
In addition, we also compare with the inference-stage debiasing approach proposed by \cite{tang2023found} to investigate the complementary effect between inference-stage and tuning-stage debiasing in LLM-listwise reranking. RankZephyr, First, and our method employ a sliding window approach using a window size of 20 and a step size of 10.

\header{Training Configuration} Following our baselines \cite{pradeep2023rankzephyr,reddy2024firstfasterimprovedlistwise}, we use Zephyr\(\beta\) \cite{tunstall2023zephyr}, an instruction-following 7-billion parameter LLM based on Mistral \cite{jiang2023mistral}, for listwise reranking. We fine-tune Zephyr for listwise reranking for three epochs, using an effective batch size of 8 with gradient accumulation of 4, a learning rate of \(5e{-6}\), and bf16. Additionally, we integrate noisy embeddings \cite{jain2023neftune} to enhance robustness. Training takes about 7 hours on four 40GB Nvidia A100 GPUs using DeepSpeed \cite{rasley2020deepspeed}. We use \(\lambda = 0.1\) for scaling the weighted RankNet loss.

\header{Propensity estimation} We sampled 3,000 queries from the MS MARCO training set.  For each query, we retrieved the top 20 candidate passages via BM25. These 20 passages were shuffled 10 times per query, resulting in a total of 30,000 samples for propensity estimation. We employed the First model checkpoint released in~\cite{reddy2024firstfasterimprovedlistwise} to re-rank the above samples and estimate propensities.

\begin{table*}[ht]
    \caption{Evaluation of positional calibration with IPS and Pos-Aug on in-domain (TREC and MS MARCO) and out-of-domain (BEIR) Datasets; All reranking is conducted using Contriever as the first-stage retriever; \textit{\(\dagger\) \(\ddagger\)  indicates a paired significant t-test \(p < 0.01\).  (\(\dagger\) indicates a test when compared to First. \(\ddagger\) indicates a test when compared to RankZephyr.); We used the following abbreviations for dataset names: HotpotQA (HQA), NFCorpus (NFC), DBPedia (DBP), Trec-covid (Tcovid), Climate-Fever (CFever), and MS MARCO (MSM).}} 
    \label{tab: table 1}
    \centering
    \renewcommand{\arraystretch}{1.4} 
     \resizebox{\linewidth}{!}{%
    \begin{tabular}{cccclllllcllcll}
    \toprule
    \toprule

 & & \multicolumn{2}{c}{\textbf{TREC}}& \multicolumn{11}{c}{\textbf{BEIR}}\\
   \cmidrule(lr){3-4} \cmidrule(lr){5-15}

         \textbf{Method}& \textbf{Order}&  \textbf{DL19}&  \textbf{DL20} &\textbf{MSM}&   \textbf{FiQA}& \textbf{HQA}& \textbf{NFC}&\textbf{NQ}&\textbf{Scidocs}&\textbf{Scifact}& \textbf{DBP}&\textbf{Tcovid} & \textbf{CFever}
  &\textbf{Avg.}\\
   \cmidrule(lr){3-4} \cmidrule(lr){5-15}

         ListT5(r=2) &  Original&	69.9&	67.3 &\textbf{46.3} &	41.7&	71.5&	36.3&	53.1&	17.2&	74.1&	46.1&	\textbf{80.3} &	24.4
  &49.1\\

        RankZephyr&  Original  &  69.3&  71.2 &42.7&  42.2&  71.6&  37.7&  65.6& 20.5& \textbf{76.7}& 50.0& 78.4& 25.6
  &51.1\\
        First & Original  &  68.2&  70.2 &44.3&   42.4&   74.2&  37.4&  66.3&  20.5&  74.6&  50.8&  79.0& \textbf{26.9} &51.6\\
      DebiasFirst& Original  & \textbf{70.0}& \textbf{72.0}$^\dagger$  &43.7& \textbf{44.3}$^{\dagger\ddagger}$ & \textbf{75.8}$^{\dagger\ddagger}$ & \textbf{37.8} & \textbf{68.2}$^{\dagger\ddagger}$ & \textbf{21.3}$^{\dagger\ddagger}$ & 76.6 & \textbf{51.9}$^{\dagger\ddagger}$ & 79.6 & 24.9
  &\textbf{52.4}\\

   \cmidrule(lr){3-4} \cmidrule(lr){5-15}

 ListT5(r=2)& Shuffle&	69.0&	67.9 &\textbf{45.8}&	42.0&	70.0&	35.7&	57.9&	18.0&	73.9&	44.3&	77.0&	23.8
  &48.8\\
 RankZephyr& Shuffle& 69.6& 70.7 &41.0& 39.9& 70.1& 37.2& 65.1& 19.4& 74.3& 49.1& 76.9& \textbf{25.4}
  &49.8\\

     First model&  Shuffle  &  67.3&  69.0 &42.1&   39.2&   73.3&  35.1&  65.6&  18.9&  73.5&  48.6&  76.4& 25.4
  &49.8\\
DebiasFirst&  Shuffle&  \textbf{71.1}$^\dagger$ &  \textbf{71.4} &43.4$^\ddagger$ &  \textbf{44.3}$^{\dagger\ddagger}$&  \textbf{75.5}$^{\dagger\ddagger}$&  \textbf{37.6}$^\dagger$&  \textbf{68.1}$^{\dagger\ddagger}$& \textbf{20.7}$^{\dagger\ddagger}$& \textbf{75.5}& \textbf{51.7}$^{\dagger\ddagger}$& \textbf{78.8}& 24.6
 &\textbf{52.0}\\
\bottomrule
\bottomrule
    \end{tabular} 
    } 
\end{table*}

\section{Results and Discussion}
In this section, we evaluate the effectiveness of our method in mitigating positional bias through a controlled position experiment on the MS MARCO development set (RQ1). We then examine whether reducing positional bias improves reranking performance across diverse datasets (RQ2) and under different first-stage retrievers (RQ3). Finally, we analyze the complementary benefits of combining our approach with inference-stage debiasing methods (RQ4).

\noindent
\textbf{RQ1: How effective is DebiasFirst in reducing positional bias?} We conduct a controlled positional bias analysis using the MS MARCO dev set. Specifically, we select the top 20 candidate passages retrieved by the first-stage retriever Contriever \cite{izacard2021contriever} and systematically place the relevant passage at different positions (from 1 to 20). By varying the position of the relevant passage, we are able to directly assess the efficacy of our method in handling changes to the relevant passage in the input order. 
As illustrated in Figure~\ref{fig: baselines}, DebiasFirst maintains remarkably consistent performance across all positions, while the baselines falter when the relevant passage appears later in the list. This stability of DebiasFirst  highlights its ability to reduce reliance on initial passage placement.

However, the positional bias mitigation in LLM-based listwise reranking introduces trade-offs. By distributing attention evenly across positions, DebiasFirst underperforms RankZephyr~\cite{pradeep2023rankzephyr} in the first three positions and trails both First and ListT5~\cite{yoon2024listt5} in the first four. This shift reflects a focus on equitable treatment of all input positions, slightly reducing emphasis on the front positions of input. These observations prompt further analysis of DebiasFirst’s performance across diverse datasets (RQ2) and with different first-stage retrievers (RQ3).

\begin{table*}
    \centering
    \caption{Evaluation of the robustness with different first-stage retriever; All models are evaluated using NDCG@10.  
    \textit{A \(\dagger\) indicates a paired significant t-test \(p < 0.01\).  \(\dagger\) indicates a test when compared to First with same first-stage retriever.}}
    \renewcommand{\arraystretch}{1.3} %
    \resizebox{\linewidth}{!}{%
    \begin{tabular}{cccclccclllllll}
    \toprule
    \toprule
 & & \multicolumn{2}{c}{\textbf{TREC}}& \multicolumn{11}{c}{\textbf{BEIR}}\\
   \cmidrule(lr){3-4} \cmidrule(lr){5-15}
   
   \textbf{Reranker}&  \textbf{First-stage}&  \textbf{DL19}&  \textbf{DL20} &\textbf{MSM}&  \textbf{FiQA}&  \textbf{HQA}&  \textbf{NFC}& \textbf{NQ}& \textbf{Scidocs}& \textbf{Scifact}& \textbf{DBP}& \textbf{Tcovid}& \textbf{CFever} &\textbf{Avg.}\\

   \cmidrule(lr){3-4} \cmidrule(lr){5-15}

 & &\multicolumn{13}{c}{\textbf{First-stage Retriever Effectiveness}}\\
    \cmidrule(lr){3-4} \cmidrule(lr){5-15}
           -&  BM25&  50.6&  48.0 &22.8&  23.6&  63.3&  32.2& 30.6& 14.9& 67.9& 31.8& 59.5& 16.5 &37.8\\
           -&  Contriever&  44.5&  42.1 &40.7&  32.9&  63.8&  32.8& 49.8& 16.5& 67.7& 41.3& 59.6& 23.7 &43.1\\
           -&  Splade++&  73.2&  72.0 &44.9&  34.8&  68.7&  34.7& 53.8& 15.9& 70.4& 43.7& 72.7& 23.0 &46.4\\
 -& RRF& 66.8& 61.0 &39.1& 34.5& 69.0& 35.2& 50.0& 17.4& 72.9& 44.4& 78.0&26.3 &47.5\\
   \cmidrule(lr){3-4} \cmidrule(lr){5-15}

    &  &\multicolumn{13}{c}{\textbf{Second-stage Reranker Effectiveness}}\\
   \cmidrule(lr){3-4} \cmidrule(lr){5-15}

   First&  BM25&  72.7&  71.1 &37.7&  39.3&  74.7&  \textbf{32.2}& 57.5& 19.5& 75.4& 46.2& 83.1& \textbf{24.0} &49.0\\
   DebiasFirst&  BM25&  \textbf{74.5}&  \textbf{72.7} &\textbf{38.5}&  \textbf{41.4}$^{\dagger}$&  \textbf{76.3}$^{\dagger}$&  \textbf{32.2}& \textbf{59.1}$^{\dagger}$& \textbf{20.7}$^{\dagger}$& \textbf{76.3}& \textbf{46.7}& \textbf{86.1}$^{\dagger}$ & 23.7 &\textbf{50.1}\\
   
   First&  Contriever&  68.2&  70.2 &\textbf{44.3}&  42.4&  74.2&  37.4& 66.3& 20.5& 74.6& 50.8& 79.0& \textbf{26.9} &51.6\\
 
  DebiasFirst& Contriever  & \textbf{70.0}& \textbf{72.0}$^\dagger$  &43.7& \textbf{44.3}$^{\dagger}$ & \textbf{75.8}$^{\dagger}$ & \textbf{37.8} & \textbf{68.2}$^{\dagger}$ & \textbf{21.3}$^{\dagger}$ & \textbf{76.6} & \textbf{51.9}$^{\dagger}$ & \textbf{79.6} & 24.9&\textbf{52.4}\\

 First& Splade++& 75.6& 79.4 &\textbf{45.1}& 42.8& 76.7& 37.5& 66.6& 20.0& 75.7& 51.3& 85.6& \textbf{26.3} &52.7\\
 DebiasFirst& Splade++& \textbf{76.9}& \textbf{82.2}$^{\dagger}$ &43.9& \textbf{44.4}$^{\dagger}$& \textbf{78.3}$^{\dagger}$& 37.5& \textbf{68.5}$^{\dagger}$& \textbf{21.3}$^{\dagger}$& \textbf{76.0}& \textbf{52.3}$^{\dagger}$& \textbf{86.6}&24.6 &\textbf{53.3}\\
 
 First& RRF& 77.3& 80.0 &\textbf{44.2}& 42.7& 77.1& 37.8& 67.0& 20.2& 76.1& 52.5& 87.1& \textbf{26.6} &53.1\\
 DebiasFirst& RRF& \textbf{78.4}& \textbf{82.4}$^{\dagger}$ &44.0& \textbf{44.8}$^{\dagger}$& \textbf{78.6}& \textbf{38.7}& \textbf{68.7}$^{\dagger}$& \textbf{21.6}$^{\dagger}$& \textbf{76.2}& \textbf{53.3}$^{\dagger}$& \textbf{88.0}$^{\dagger}$&24.7 &\textbf{53.9}\\
 \bottomrule
 \bottomrule
    \end{tabular}
    }
    \label{tab: table 2}
\end{table*}

\noindent
\textbf{RQ2: Does the elimination of positional bias help to improve the performance of reranking across datasets?} To explore this, we evaluated DebiasFirst and baselines under two conditions: (1) the original order, where passages are sorted by the first-stage retriever, typically placing relevant passages at the top; and (2) a shuffled order, where relevant passages are placed in a random position of input, as shown in Table \ref{tab: table 1}. The shuffled order evaluation tries to mirror real-world challenges, such as dynamic news ranking aggregator~\cite{Chakraborty_2017_Optimizing,Styskin_2011}, where articles stream in by recency rather than relevance, and federated search~\cite{Shokouhi2009}, where passages from multiple sources merge without a unified relevance score. All reranking is conducted using Contriever as the first-stage retriever.

\header{Impact on original and shuffled order} In the original order, DebiasFirst outperforms the baselines (First, RankZephyr, and ListT5) across both in-domain and out-of-domain datasets. It significantly surpasses First and RankZephyr in 6 of 12 datasets and achieves notable gains over RankZephyr in the remaining 6. This strong performance indicates that eliminating positional bias does not reduce overall effectiveness, even when relevant passages are placed at the top. In the shuffled order, DebiasFirst maintains its effectiveness. Unlike the baselines, whose performance reduces with randomized passage order, DebiasFirst exhibits only a minimal performance drop compared to the original order. 

\begin{figure*}[h]
    \centering
    \includegraphics[width=0.5\linewidth]{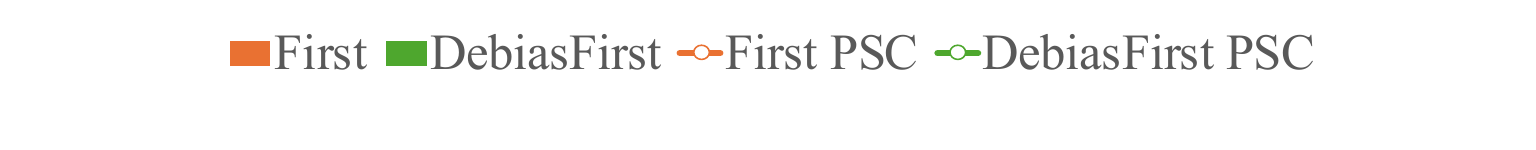}
    
    \begin{subfigure}{0.47\textwidth}
        \includegraphics[width=\linewidth]{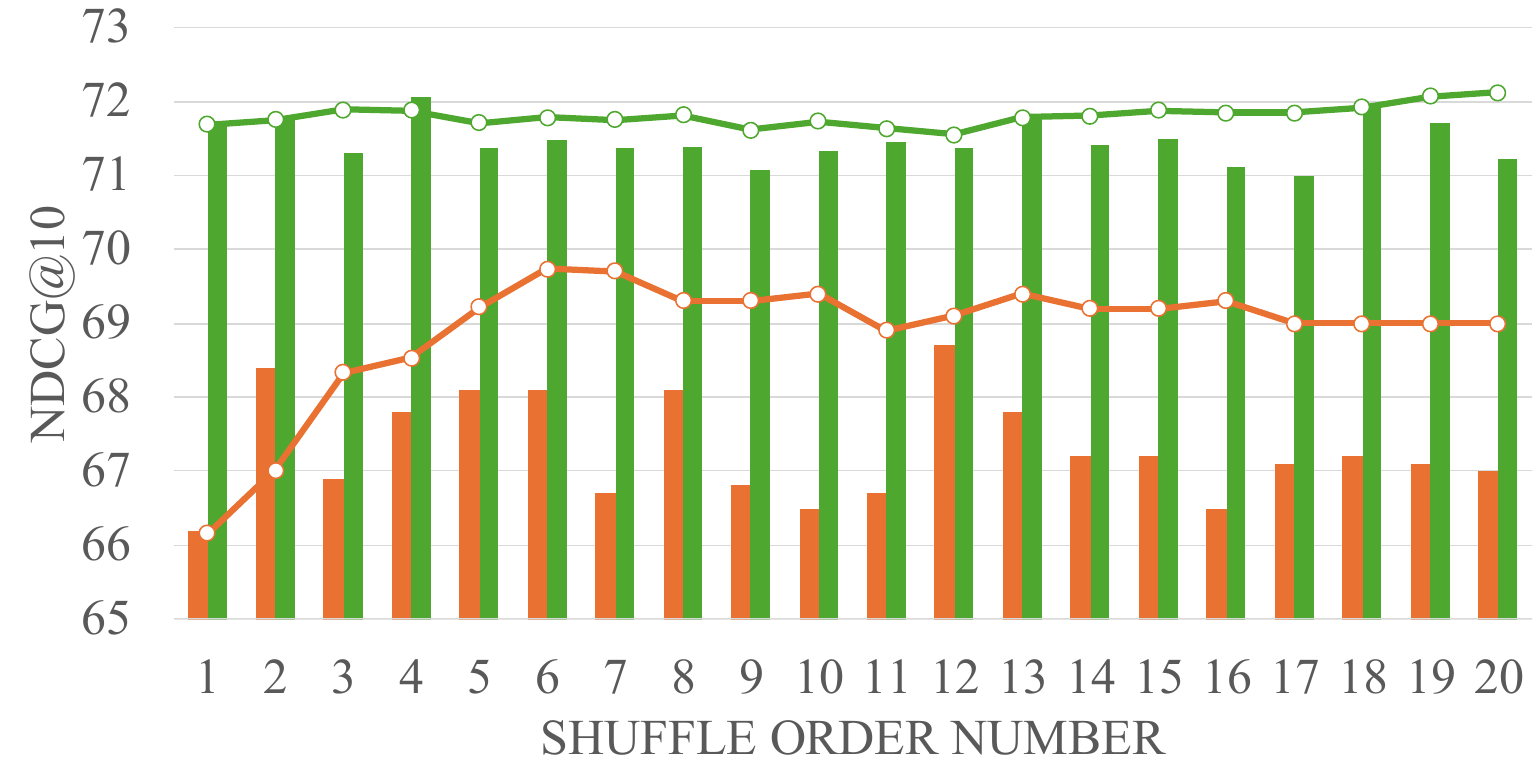}
        \caption{Evaluation on DL2019}
    \end{subfigure}
    \hfill 
    \begin{subfigure}{0.47\textwidth}
        \includegraphics[width=\linewidth]{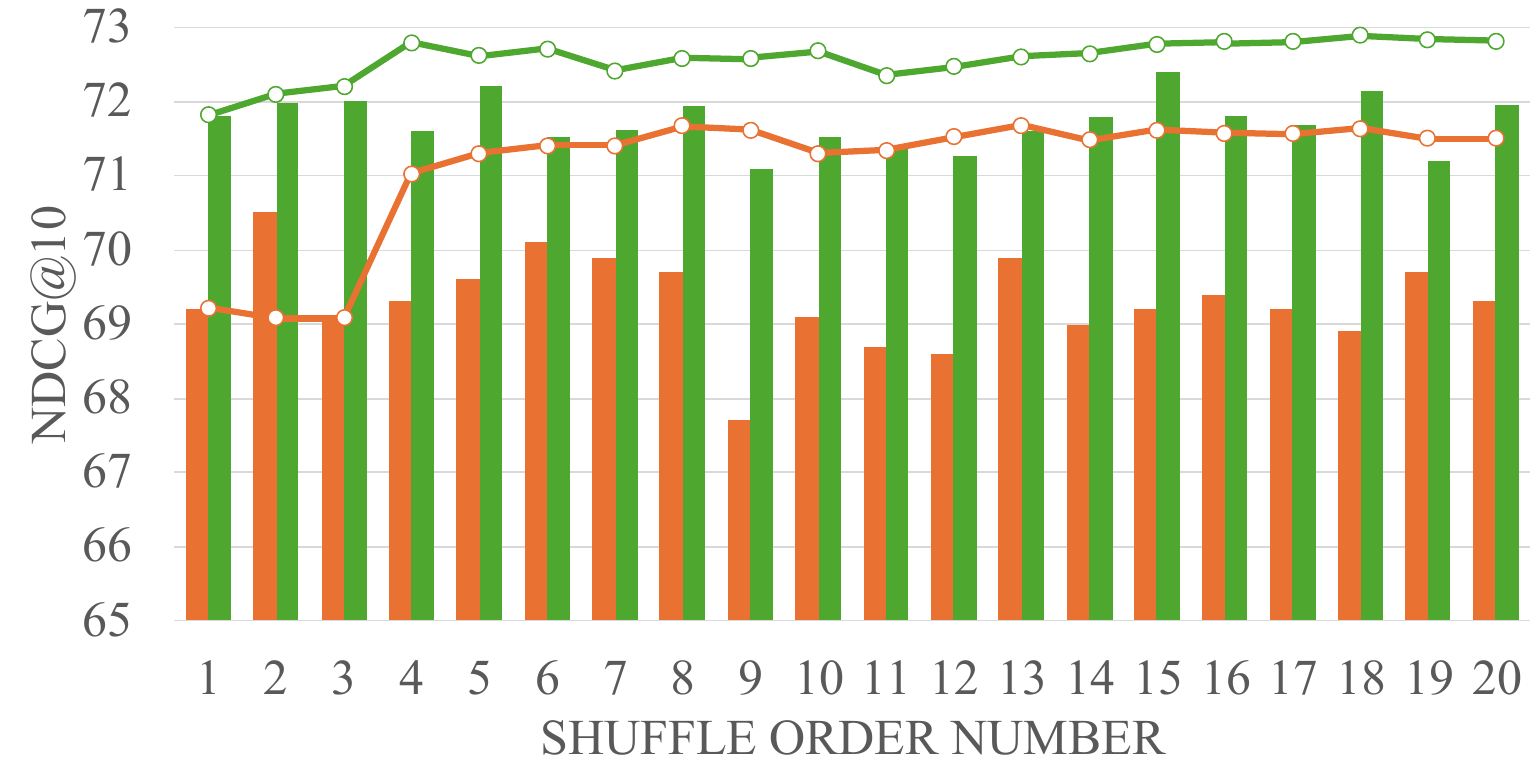}
        \caption{Evaluation on DL2020}
    \end{subfigure}
    \caption{Evaluating the complementary effects of inference-Stage (PermSC approach \cite{tang2023found}) and tuning-stage (our approach) debiasing in LLM-Based listwise passage reranking;
    \textit{Bars represent the performance on each shuffled ordering; Lines represent the aggregated performance using PermSC rank aggregation.}
    } 
    \label{fig:5}
\end{figure*}

\noindent
\textbf{RQ3: Does DebiasFirst remain effective with different first-stage retrievers?} We evaluate how DebiasFirst performs against three different first-stage retrievers: Contriever, Splade++ \cite{splade++}, and BM25. We additionally apply reciprocal rank fusion (RRF) \cite{cormack2009reciprocal} to fuse results from the three first-stage retrievers.

As shown in Table~\ref{tab: table 2}, DebiasFirst consistently outperforms the baseline First on most datasets. With weaker retrievers like BM25, DebiasFirst mitigates low-quality initial rankings by fairly assessing passage relevance in all positions. With stronger retrievers, it refines already-strong rankings. Overall, DebiasFirst demonstrates robust performance across various retrievers.

\noindent
\textbf{RQ4: Does DebiasFirst outperform the inference-stage debiasing baseline?} Previous literature PermSC~\cite{tang2023found} focuses on eliminating positional bias by aggregating rankings from multiple reranked input orders. PermSC computes a central ranking that minimizes positional bias by identifying the ranking closest to all permutations in terms of distance metrics. To evaluate whether our method surpasses the inference-stage method, we implemented PermSC~\cite{tang2023found} rank aggregation with DebiasFirst and First~\cite{reddy2024firstfasterimprovedlistwise} model. Figure~\ref{fig:5} shows the performance of DebiasFirst compared to First~\cite{reddy2024firstfasterimprovedlistwise} across 20 different shuffled inputs.  

\header{Effectiveness of inference vs. tuning stage debiasing} On DL2019, DebiasFirst (green bar) consistently outperformed the First with PermSC rank aggregation (orange line) across all 20 runs. On DL2020, DebiasFirst is inferior to First with PermSC in 3 out of 20 runs. This difference is attributed to the stable performance of DebiasFirst on DL2019, which exhibited a variance of 0.08 across all runs, compared to a higher variance of 0.120 on DL2020. Overall, the results demonstrate that debiasing during the tuning phase is more effective than implementing it during the inference phase.

\header{Complementary benefits of inference-stage debiasing} We also investigated whether applying PermSC rank aggregation at the inference stage remains beneficial for models already debiased during tuning. To this end, we implemented PermSC rank aggregation with both DebiasFirst and First, and analyzed their performance across 20 shuffled runs on DL2019 and DL2020. Results show PermSC rank aggregation (green line) can still improve performance of DebiasFirst (green bar), but the improvement becomes less. Furthermore, DebiasFirst achieves optimal performance with fewer shuffles than First. It reaches peak performance at shuffle order 4 for DL2019 and 6 for DL2020, compared to orders 6 and 8 for First, respectively. Thus, while PermSC rank aggregation can still provide a complementary benefit, it is less critical for DebiasFirst, as its tuning-stage debiasing already makes it robust to input order. 

\begin{figure*}[h]
    \centering
    \begin{subfigure}{0.48\textwidth}
        \includegraphics[width=\linewidth]{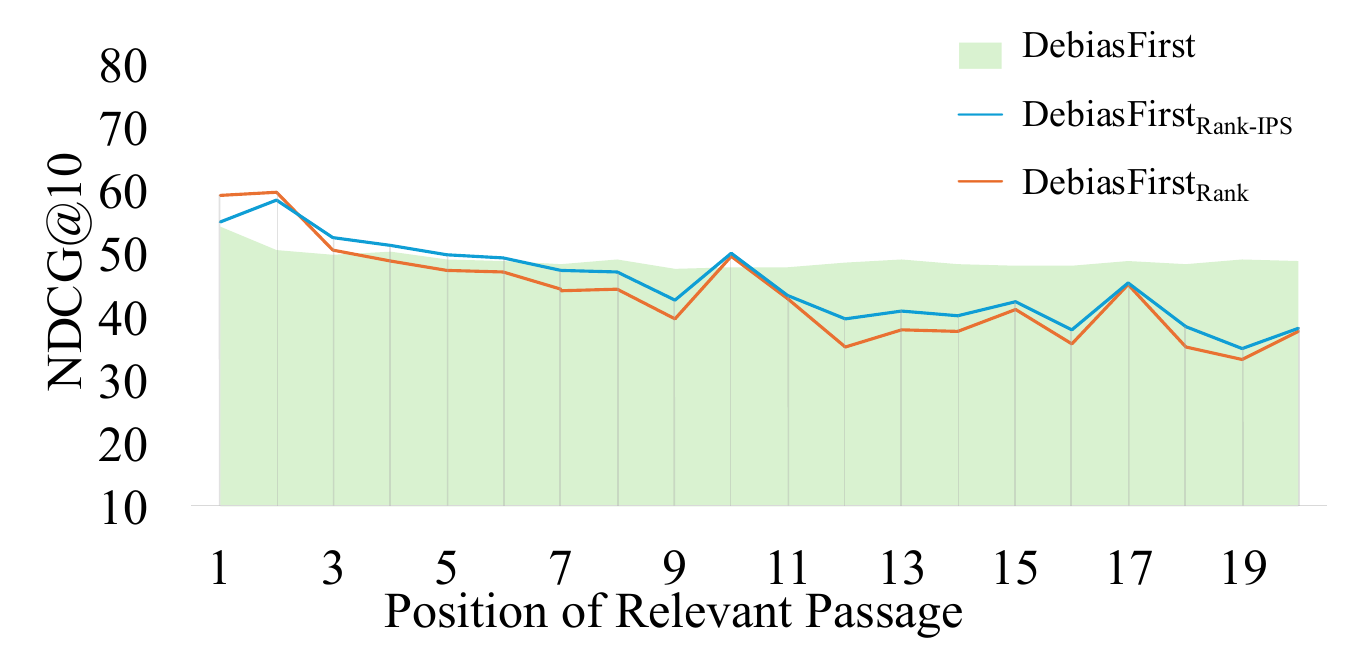}
        \caption{Comparison of IPS Strategy w/o LM}
    \end{subfigure}
    \hfill 
    \begin{subfigure}{0.48\textwidth}
        \includegraphics[width=\linewidth]{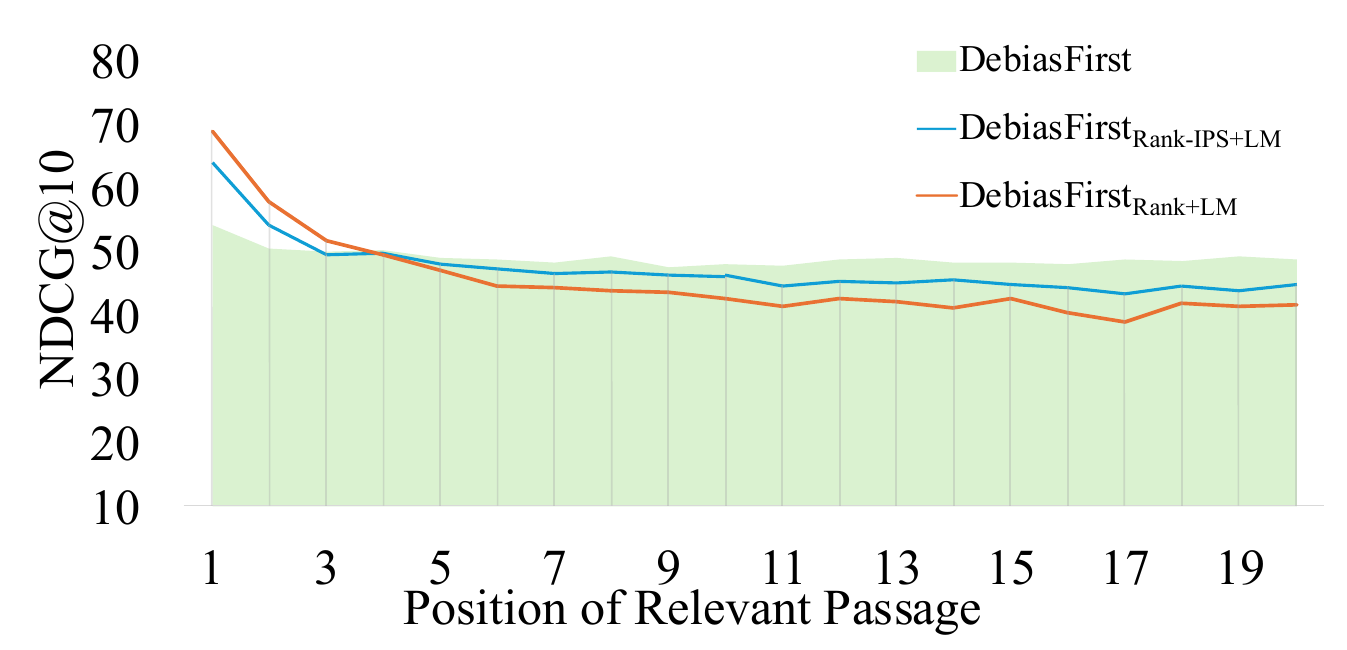}
        \caption{Comparison of IPS Strategy with LM}
    \end{subfigure}
    \hfill 
    \begin{subfigure}{0.48\textwidth}
        \includegraphics[width=\linewidth]{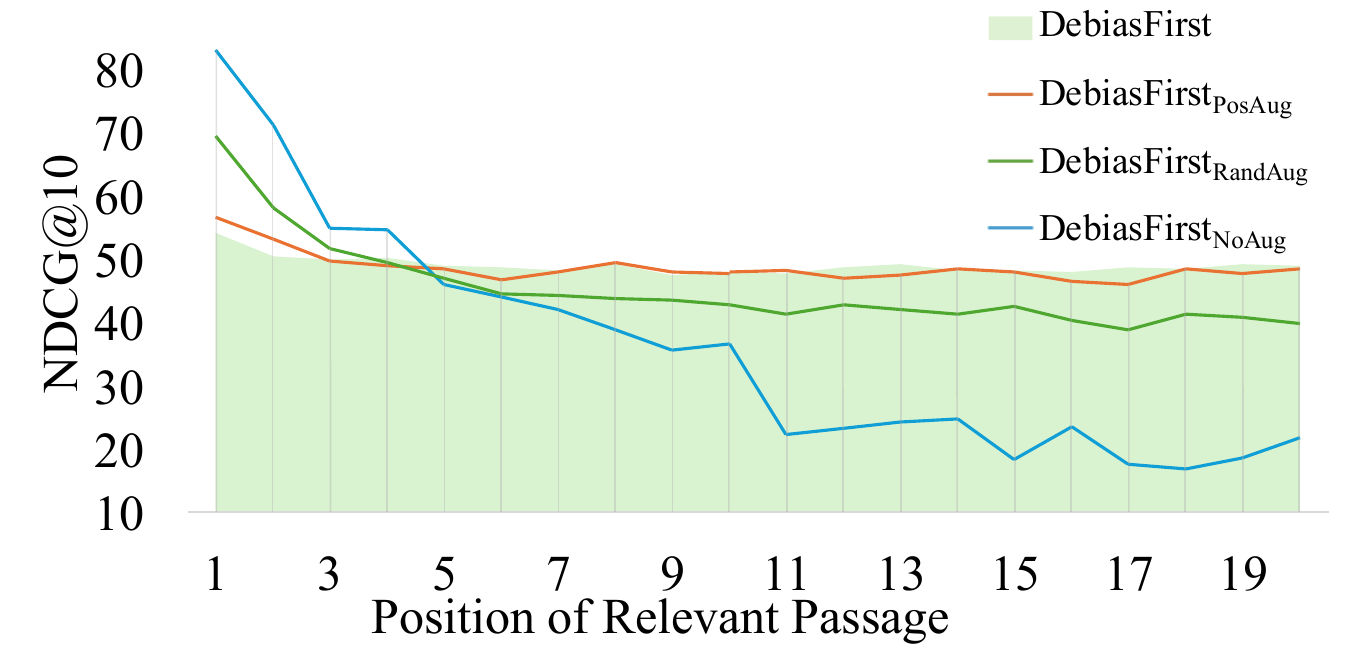}
        \caption{Comparison of Augmentation Strategy}
    \end{subfigure}
    \hfill 
    \begin{subfigure}{0.48\textwidth}
        \includegraphics[width=\linewidth]{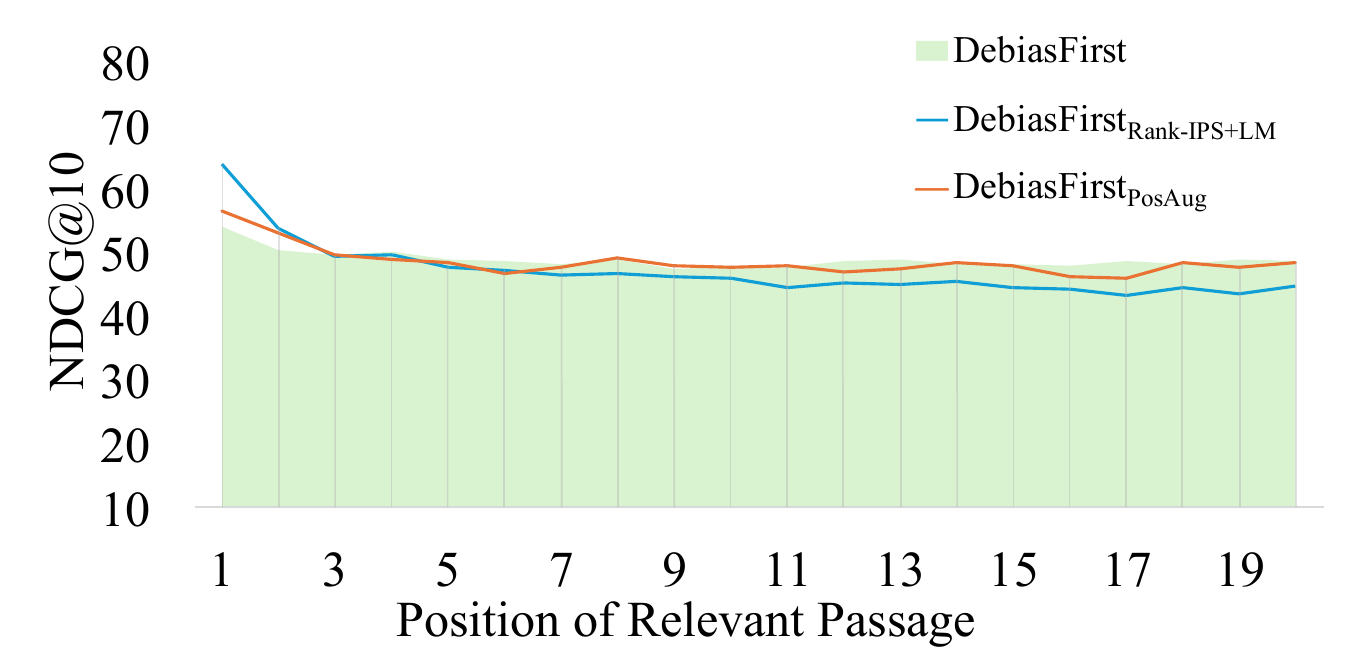}
        \caption{Comparison of Synergistic Impact}
    \end{subfigure}
    \caption{Comparison of positional calibration with IPS and Pos-Aug on the MS MARCO Dev Set. All models are evaluated using NDCG@10. }
    \label{fig: ablation_study_control_experiment}
\end{figure*}

\section{Ablation Study}
To assess each component's effectiveness in reducing positional bias, we performed an ablation study by isolating key elements and evaluating their effects on ranking performance. We tested five variants: (1) DebiasFirst$_{\textrm{NoAug}}$, which excludes any augmentation; (2) DebiasFirst$_{\textrm{RandAug}}$ using the random augmentation strategy used by RankZephyr~\cite{pradeep2023rankzephyr}; (3) DebiasFirst$_{\textrm{Rank}}$ excluding both IPS calibration and LM optimization objective; (4) DebiasFirst$_{\textrm{Rank-IPS}}$ using IPS calibration but excluding LM loss; and (5) DebiasFirst$_{\textrm{Rank-IPS+LM}}$ incorporating both IPS calibration and LM optimization objective.

\header{Impact of IPS} We compared variants with and without IPS across all 20 input positions, as in Figure~\ref{fig: ablation_study_control_experiment} (a-b). Results demonstrate that IPS calibration reduces performance variance across positions. Compared to the baseline variant (\textrm{DebiasFirst}$_{\textrm{Rank}}$), the IPS-enhanced model (\textrm{DebiasFirst}$_{\textrm{Rank-IPS}}$) typically achieves equal or superior performance. Integrating a LM optimization objective further enhances performance and reduces positional variance. 

\begin{table*}[ht]
    \caption{Evaluation of positional calibration with IPS and Pos-Aug on in-domain (TREC and MS MARCO) and out-of-domain (BEIR) Datasets.  \textit{A \(\dagger\)   indicates a paired significant t-test \(p < 0.01\).  (\(\dagger\) indicates a test when compared to DebiasFirst$_{RandAug}$ without IPS calibration.)}}
    \label{tab: table 3}
    \centering
    \renewcommand{\arraystretch}{1.3} 
    \resizebox{\linewidth}{!}{%
    \begin{tabular}{cccclllllcllcll}
    \toprule
    \toprule
 & & \multicolumn{2}{c}{\textbf{TREC}}& \multicolumn{11}{c}{\textbf{BEIR}}\\
 \cmidrule(lr){3-4} \cmidrule(lr){5-15}
         \textbf{Strategy}& \textbf{Order}&  \textbf{DL19}&  \textbf{DL20} &\textbf{MSM}&   \textbf{FiQA}& \textbf{HQA}& \textbf{NFC}&\textbf{NQ}&\textbf{Scidocs}&\textbf{Scifact}& \textbf{DBP}&\textbf{Tcovid} & \textbf{CFever}
  &\textbf{Avg.}\\
   \cmidrule(lr){3-4} \cmidrule(lr){5-15}

        DebiasFirst$_{NoAug}$&  Original  &  57.7&  54.9 &43.5&   40.7&   74.6&  35.9&  65.2&  19.1&  75.9&  46.7&  72.0& 
 \textbf{27.9} &50.2\\

        DebiasFirst$_{RandAug}$&  Original  &  68.5&  70.2 &\textbf{44.5}&   42.4&   74.2&  37.4&  66.1&  20.4&  74.6&  50.8&  79.1& 
 26.7 &51.6\\
        
        DebiasFirst$_{PosAug}$&  Original  &  69.7&  70.6 &43.9&  44.2$^\dagger$ &  75.7$^{\dagger}$ &  37.3&  68.1$^{\dagger}$& \textbf{21.4}$^\dagger$& 74.8& 51.4& 78.6& 23.7
  &51.9\\
         \textrm{DebiasFirst}$_{\textrm{Rank-IPS+LM}}$ & Original  & \textbf{70.2} & 71.2 &44.4 & \textbf{44.6}$^{\dagger}$ & 75.5$^{\dagger}$ & \textbf{37.8} & \textbf{68.7}$^{\dagger}$ & 20.7& \textbf{76.9}$^\dagger$ & \textbf{51.9}$^{\dagger}$ & 78.4& 25.6 &\textbf{52.5}\\
          DebiasFirst& Original  & 70.0& \textbf{72.0}$^\dagger$  &43.7& 44.3$^{\dagger}$ & \textbf{75.8}$^{\dagger}$ & 37.8 & 68.2$^{\dagger}$ & 21.3$^{\dagger}$ & 76.6 & \textbf{51.9}$^{\dagger}$ & 79.6 & 24.9
  &52.4\\
 \cmidrule(lr){3-4} \cmidrule(lr){5-15}

        DebiasFirst$_{NoAug}$&  Shuffle  &  49.0&  48.4 &27.2&   29.4&   67.0&  27.7&  55.1&  13.0&  66.3&  35.4&  60.0& 
 22.0 &40.3\\
        DebiasFirst$_{RandAug}$& Shuffle  &  67.3&  69.0 &42.1&   39.2&   73.3&  
        35.1&  65.6&  18.9&  73.5&  48.6&  76.4& \textbf{25.4} &49.8\\

DebiasFirst$_{PosAug}$& Shuffle& 69.3& 70.3 &43.3$^\dagger$ & 43.7$^\dagger$ & 75.4$^{\dagger}$& 37.7$^\dagger$ & 67.8$^{\dagger}$ & \textbf{21.2}$^{\dagger}$ & 75.1& 51.3$^\dagger$ & 78.7& 24.5
 &51.9\\

\textrm{DebiasFirst}$_{\textrm{Rank-IPS+LM}}$ &Shuffle& 69.6& 71.2 &\textbf{43.4}& \textbf{44.6}$^\dagger$ & 75.3$^\dagger$ & 37.0$^\dagger$ &\textbf{68.5}$^\dagger$ & 20.1$^\dagger$ &\textbf{75.7}$^\dagger$ & 51.5$^\dagger$ & 78.0& 25.2 &\textbf{51.9}\\
DebiasFirst&  Shuffle&  \textbf{71.1}$^\dagger$ &  \textbf{71.4}  &\textbf{43.4} &  44.3$^{\dagger}$ &  \textbf{75.5}$^{\dagger}$&  \textbf{37.6}$^\dagger$ &  68.1$^{\dagger}$ & 20.7$^{\dagger}$ & 75.5& \textbf{51.7}$^{\dagger}$ & \textbf{78.8}& 24.6
 &52.0\\
\bottomrule
\bottomrule
    \end{tabular}
    }
\end{table*}

\header{Impact of position-aware augmentation} To evaluate the effectiveness of position-aware augmentation, we compared three variants: DebiasFirst$_{\textrm{NoAug}}$, DebiasFirst$_{\textrm{RandAug}}$ and \textrm{DebiasFirst}$_{\textrm{PosAug}}$ (Figure~\ref{fig: ablation_study_control_experiment} (c)). The variant without augmentation (blue line) exhibits a significant performance decline from the first to the last input position, indicating a high degree of positional bias. Random augmentation (green line) helps reduces this drop, but it does not fully mitigate the positional bias issue. In contrast, \textrm{DebiasFirst}${\textrm{PosAug}}$, which more evenly distributes training instances across input and output positions, demonstrated the lowest performance variance across input positions. This result suggests that more evenly distributing training instances across input and output positions can serve as an effective calibration strategy.

\header{Synergistic Impact} The combined effect of IPS and Pos-Aug is indicated by the green squares in Figure~\ref{fig: ablation_study_control_experiment}. This combined variant further reduces variance across input positions and leads to a more stable performance across all input positions. It consistently outperforms individual configuration.

\header{Impact on overall ranking performance} We finally evaluated the overall ranking performance of each variant on both MS MARCO and BEIR datasets in Table~\ref{tab: table 3}. The Pos-Aug augmentation variants, \textrm{DebiasFirst}$_{\textrm{PosAug}}$, showed substantial improvements, outperforming \textrm{DebiasFirst}$_{\textrm{RandAug}}$ in 9 out of 12 datasets on original order, and 10 out of 12 datasets on shuffled order. Similarly, the \textrm{DebiasFirst}$_{\textrm{Rank-IPS+LM}}$ demonstrates a clear advantage compared to \textrm{DebiasFirst}$_{\textrm{RandAug}}$ without IPS calibration, outperforming it in 9 out of 12 datasets on original order and 11 out of 12 datasets on shuffled order. Finally, the full model DebiasFirst, which integrates both IPS and Pos-Aug, achieves a better overall performance compared to all other variants. 

\vspace{-2mm}
\section{Related Work}
\header{LLMs as Listwise Rankers} Recent studies~\cite{sun2023chatgptgoodsearchinvestigating,pradeep2023rankvicunazeroshotlistwisedocument,pradeep2023rankzephyr,reddy2024firstfasterimprovedlistwise,yoon2024listt5,liu2024slidingwindowsendexploring,Ruiyang_selfcalibrated_2025} have leveraged LLMs to simultaneously rank lists of passages by producing reranked document identifiers. These methods can be broadly categorized into two approaches: one that predicts the entire sequence of passage identifiers~\cite{sun2023chatgptgoodsearchinvestigating,pradeep2023rankvicunazeroshotlistwisedocument,pradeep2023rankzephyr,yoon2024listt5}, and another that generates a single token and utilizes its output logits for ranking~\cite{reddy2024firstfasterimprovedlistwise,Zhuang_2024}. The single token approach is much more efficient than the full token approach, as it only needs to be decoded once for each ranking process. In this study, we choose to build on the efficient single-token generation approach for listwise passage reranking.

\header{Mitigating positional bias by output aggregations} Despite the success of LLMs in listwise reranking, their effectiveness is adversely affected by positional bias~\cite{tang2023found}. To address this, some studies mitigate bias at inference through output aggregation~\cite{hou2024largelanguagemodelszeroshot,zeng2024llmrankfusionmitigatingintrinsicinconsistency,wang2023large,li2024splitmergealigningposition}. They aim to improve output consistency by aggregating rankings derived from multiple runs with varied candidate orders. Specifically, Tang et al.~\cite{tang2023found} aggregate multiple rankings by minimizing the Kendall tau distance across all sampled rankings. LLM-RankFusion~\cite{zeng2024llmrankfusionmitigatingintrinsicinconsistency} enhances order consistencies by employing in-context learning for order-agnostic comparisons and calibrating preference probabilities. 
However, inference-stage bias mitigation methods significantly increase computational demands, posing challenges for practical implementations in real-time or large-scale systems. 

\header{Mitigate positional bias by fine-tuning} Other studies~\cite{pradeep2023rankvicuna,pradeep2023rankzephyr}, have attempted to mitigate positional bias through random shuffling augmentation, but have found that it sacrifices overall effectiveness. Alternatively, Yoon et al.~\cite{yoon2024listt5} tackled positional bias by employing tournament sort to predict from least to most relevant, but at the cost of increasing computational overload. 
In unbiased learning-to-rank, inverse propensity scoring (IPS) is a widely used approach to mitigate bias in user clicks~\cite{Joachims2017ublr,oosterhuis2021unifying,wang2016,Ai_2018,hager2024unbiased}. 
Drawing on this, we propose fine-tuning LLMs with IPS to estimate and correct positional bias, improving listwise reranking effectiveness.
\vspace{-3mm}
\section{Conclusion}
In this study, we introduce DebiasFirst, a method designed to mitigate positional bias for listwise reranking by integrating positional calibration with inverse propensity scoring (IPS) and position-aware augmentation. We show that both positional calibration and position-aware augmentation effectively reduce positional bias, particularly in enhancing the ranking performance of relevant passages positioned at the end of the input list. DebiasFirst consistently enhances ranking performance across both in-domain and out-of-domain datasets, demonstrating strong generalizability and effectiveness across diverse first-stage retrievers. However, this study is limited to reranking scenarios with a window size of 20 candidate passages. The effectiveness of IPS and Pos-Aug in handling longer context remains unexplored, such as reranking with a window size of 100 passages. Future work will focus on assessing the effectiveness of IPS and Pos-Aug in long-context settings. 
 
\subsection*{Acknowledgments} We thank all reviewers for their feedback. This research was supported by the project VI.Vidi.223.166 of the NWO Talent Programme (partly) financed by the Dutch Research Council (NWO). The views expressed in this paper are those of the authors and do not necessarily reflect the views of their institutions or sponsors. 

\subsection*{Disclosure of Interests} The authors have no competing interests to declare that are relevant to the content of this article.
\bibliographystyle{splncs04}
\bibliography{references}
\end{document}